\documentclass[english,aps,twocolumn]{revtex4}
\usepackage[T1]{fontenc}
\usepackage[latin1]{inputenc}
\usepackage{amsmath}

\makeatletter
\usepackage{babel}
\makeatother
\begin{document}

\preprint{This line only printed with preprint option}

\title{Fermions without fermi fields}

\author{R.C. Ball}

\email{r.c.ball@warwick.ac.uk}

\homepage{http://go.warwick.ac.uk/theory}

\thanks{}

\affiliation{Department of Physics, University of Warwick, Coventry CV4 7AL, UK}

\begin{abstract}
It is shown that an arbitrary Fermion hopping hamiltonian can be mapped
into a system with no fermion fields, generalising an earlier model
by Levin \& Wen\cite{Wen1,Wenlong}. All operators in the hamiltonian
of the resulting description commute (rather than anticommute) when
acting at different sites, despite the system having excitations obeying
Fermi statistics. Whilst extra conserved degrees of freedom are introduced,
they are all locally identified in the representation obtained. The
same methods apply to Majorana (half) fermions, which for cartesian
lattices mitigate the Fermion Doubling Problem. The generality of
these results suggests that the observation of Fermion excitations
in nature does not demand that anticommuting Fermion fields be fundamental.
\end{abstract}
\maketitle

As fundamental entities, fermion fields appear in conflict with the
principle of locality: all fermion creation and annihilation operators
\emph{anti}commute no matter how far apart are the points in space
at which they act (without restriction by causal connection). This
feature is built into Quantum Field Theory through the use of anticommuting
Grassman fields, and supersymmetric string theories likewise have
explicit anticommuting coordinates. Locality is only preserved in
the Physics by conservation of fermion number, forcing the fermion
operators to appear in pairs. In recent years there has been sustained
interest in how particle statistics can be manipulated\cite{anyons},
and understanding the Fractional Quantum Hall Effect\cite{fqhe} has
widenned the appreciation that the statistics of the elementary excitations
of a system need not simply reflect the statistics of its components.
Thus we ask whether it is really necessary to put fermion fields into
physics `by hand', or whether they can always be understood as excitations
emerging from quantum systems built of operators whose action is strictly
local.

It is well known how fermions relate to hard core bosons in one dimension\cite{JWtrans}
and how they can be built out of bosons in two dimensions with attached
magnetic flux\cite{fluxfermions}, and that neither approach extends
naturally to three dimensions of space. The present letter builds
on the recent work of Levin \& Wen\cite{Wen1,Wenlong}, who showed
that particular models of pairwise fermion hopping could be represented
in terms of operators obeying locality, whilst the elementary excitations
remained strictly fermionic. Their mechanism is essentially the reverse
of how Kitaev\cite{Kitaev} showed that a particular hexagonal lattice
spin model has fermionic excitations. They showed that their mechanism
worked on simple hypercubic lattices, in two and three dimensions
explicitly, and interpret it in terms of string-net condensation\cite{Wenlong,stringnet}.

Here I show that fermion hopping on an \emph{arbitrary} graph can
be mapped into the excitations of a Hamiltonian devoid of explicit
fermionic operators, in which all operators acting at different sites
commute. The dimensionality of space which the graph might approximate
is irrelevant to the mapping, which is sensitive only to local coordination
numbers (presumed finite). This strongly suggests that quite arbitrary
fermion excitation spectra can be represented, and perhaps understood,
as arising from the excitation of systems whose fundamental operators
obey locality in their commutation properties. 

We start from a generic fermion hopping hamiltonian, \begin{equation}
H_{hop}=\sum_{i}c_{i}^{+}V_{i}c_{i}+\sum_{<ij>}c_{i}^{+}t_{ij}c_{j}\label{eq:hhop}\end{equation}
where the fermion field $c_{i}$ has standard anticommutation properties
$\left\{ c_{i},c_{j}\right\} =\left\{ c_{i}^{+},c_{j}^{+}\right\} =0$
and $\left\{ c_{i}^{+},c_{j}\right\} =\delta_{ij}$, the $V_{i}$
are simple `on-site' potentials (relative to the fermion chemical
potential) and the $t_{ij}=t_{ji}^{*}$ are simple `inter-site' hopping
matrix elements. The connectivity of the graph (or lattice) is encoded
by which elements $t_{ij}$ are non-zero, but below we will need to
explicitly restrict the sum over links $<ij>$to those cases. We will
also exploit the gauge invariance of the hamiltonian, that two models
related by $t_{jk}^{(2)}=t_{jk}^{(1)}e^{i\vartheta_{jk}}$ are equivalent
(through adjustment of phase of the $c_{k}$) provided the relative
phase factors multiply to unity around all closed loops.

We now introduce new operators $S_{ij}=-S_{ji}$ modulating the hoppings
across each link, such that these operators commute with each other
and the original fermions, and with eigenvalues $s_{ij}=-s_{ji}=\pm1$.
The hamiltonian is then generalised to \begin{multline}
H_{gauge}=\sum_{i}c_{i}^{+}V_{i}c_{i}-\sum_{<ij>}i\, S_{ij}c_{i}^{+}u_{ij}c_{j}\\
+\sum_{<ij..z>}g_{ij..z}\, S_{ij}S_{j.}..S_{.z}S_{zi}.\label{eq:htowen}\end{multline}
Here $u_{ij}=is_{ij}^{0}t_{ij}$ so that the first two terms recover
the hopping hamiltonian (\ref{eq:hhop}) for a particular set of outcomes
$s_{ij}^{0}=-s_{ji}^{0}$ of the operators $S_{ij}$. We add extra
couplings $g_{ij..z}$ to (products round) `Wilson loops' of the $S_{ij}$
operators; setting $g_{ij..z}/\left(s_{ij}^{0}s_{j.}^{0}..s_{.z}^{0}s_{zi}^{0}\right)$
sufficiently negative ensures that only combinations of the $S_{ij}$
eigenvalues which are gauge transformations of the original hopping
hamiltonian contribute to the low energy states of the new hamiltonian.
Because the operators $S_{ij}$ commute with each other and the hamiltonian,
they could be separately diagonalised to their eigenvalues $s_{ij}$,
which are constants of the motion, and we then recover the original
tight binding hamiltonian (up to gauge symmetry). Thus the system
does still have its original fermion excitations.

The key to obtaining a representation with locality is now to factorise
each link operator into a pair of Majorana {[}half-{]} fermions,\[
S_{jk}=i\, m_{jk}m_{kj}\]
where the Majorana half-fermion operators are Hermitian (for simplicity)
and have anticommutation properties $\left\{ m_{jk},m_{j',k'}\right\} =2\delta_{jj'}\delta_{kk'}$
as well as \emph{anti}commuting with all the original standard fermion
operators $c_{i}$. We then associate each new half-fermion with the
site of its first index, motivating us to rewrite the hopping terms
in the hamiltonian as $b_{ij}^{+}t_{ij}b_{ji}$, where \[
b_{ij}=m_{ij}c_{i}.\]

Crucially the new operators $b_{ij}$ commute, $\left[b_{ij},b_{i'j'}\right]=0$
and $\left[b_{ij}^{+},b_{i'j'}\right]=0$, when their left (or site)
indices are unequal, $i\neq i'$. The loop terms can also be expressed
in terms of operators conforming to locality in this way. Regrouping
the factors in each loop gives us $S_{ij}S_{jk}S_{k.}..S_{.z}S_{zi}=B_{i,zj}B_{j,ik}B_{k,j.}..B_{z,.i}$
where\[
B_{j,ik}=i\, m_{ji}m_{jk}.\]
 We can finally eliminate all fermionic notation by writing $n_{i}=c_{i}^{+}c_{i}$
and hence express the hamiltonian as \begin{multline}
H_{gauge}=\sum_{i}V_{i}n_{i}+\sum_{<ij>}b_{ij}^{+}u_{ij}b_{ji}\\
+\sum_{<ij..z>}g_{ij..z}\, B_{i,zj}B_{j,i.}..B_{z,.i}.\label{eq:hwen}\end{multline}

From their definitions it is trivial to check that any pair drawn
from all the operators $n_{i}$, $b_{ij}$, $b_{ij}^{+}$, $B_{i,jk}$
appearing in the hamiltonian (\ref{eq:hwen}) commute when their first
indices are distinct. As a result we can factor the overall Hilbert
space (of wavefunctions) on which they act into a product of single
site Hilbert spaces, and the only non-trivial action of each operator
is within the corresponding single site Hilbert space. 

We now focus on each site separately and note how the required commutation
properties can be explicitly constructed. First note that each Dirac
fermion can be split into a pair of hermitian Majorana half-fermions,\begin{equation}
2c_{k}=m_{k0}+i\, m_{k\,-\!1},\label{eq:decomp}\end{equation}
in terms of which $n_{k}=\left(im_{k0}m_{k\,-\!1}+1\right)/2$. The
complete set of operator properties then required on site $k$ are
now just the Majorana anticommutation relations\[
\left\{ m_{km},\, m_{kn}\right\} =2\delta_{mn},\quad-1\le m,\, n\le z_{k},\]
where $z_{k}$ is the coordination number of (i.e. number of links
to) site $k$. These relations are obeyed by standard (Euclidean)
$4\times4$ Dirac matrices for $2+z_{k}\le5$ and by their $2^{s}\times2^{s}$
generalisations for $2+z_{k}\le2s+1$. It is crucial that we do \emph{not}
require to represent fermion anticommutation between local fermion
operators on \emph{different} sites, because all the terms in the
hamiltonian contain an even number of fermion factors from each site.
Writing $\gamma_{k}(i)$ for the $k$'th Dirac matrix acting in the
Hilbert space of the $i$'th site, we then have all the operators
in the bosonic hamiltonian (\ref{eq:hwen}) expressed in terms of
these:\begin{multline}
n_{j}=\frac{i}{2}\gamma_{0}(j)\gamma_{-1}(j)+\frac{1}{2},\label{eq:defs}\\
b_{jk}=\frac{1}{2}\gamma_{k'}(j)\left(\gamma_{0}(j)+i\,\gamma_{-1}(j)\right),\\
B_{j,ik}=i\,\gamma_{i'}(j)\gamma_{k'}(j).\end{multline}
Here on site $j$, $1\le k'(j,k)\le z_{j}$ denotes the local index
associated with its link to site $k$. 

The end result is that all the operators appearing in the Hamiltonian
are expressed in terms of local matrix operators, which in turn are
all equivalent to bilinear combinations of Dirac matrices: these might
loosely be termed `spins'. Where the original hopping hamiltonian
had links with explicit fermion hopping, the derived generalisation
of the LW model\cite{Wen1} has coupling between the local spins.
Most importantly, the spin operators for different sites commute -
yet by construction the system still has the original fermionic excitations.

The loop products of operators $B_{i,jk}$ can be further simplified
in terms of products of new operators\begin{equation}
P_{ij}=\gamma_{j'(i,j)}(i)\gamma_{i'(j,i)}(j)\label{eq:Pdef}\end{equation}
 in direct index correspondence with the original operators $S_{ij}$.
These new operators are \emph{not} equivalent to the $S_{ij}$ operators:
in particular two operators $P_{ij}$ \emph{anti}commute if they have
one site index in common.

Locality of all operators has been gained at the expense of enlarging
the Hilbert space. The original hopping hamiltonian (\ref{eq:hhop})
acted on a Hilbert space of dimension $2^{N}$ where $N$ is the number
of sites, equivalent to $N$ qbits, one for each site. To construct
the generalised Wen Hamiltonian (\ref{eq:hwen} with \ref{eq:defs})
we first added qbits equal to the number of links, $A$. However in
the final local representation we carried fewer non-commutations,
and multiplying the dimension of all the local Hilbert spaces leads
to a dimension equivalent to total qbit count\[
C=N_{E}+\frac{1}{2}N_{O}+A.\]
Here $N_{E}$ and $N_{O}$ are the counts of sites with even and odd
coordination number respectively, the even being less efficiently
represented because the number of anticommuting Dirac matrices is
naturally odd.

We should now expect that there are $A-\frac{1}{2}N_{O}$ constants
of the motion, and these and one more can be found as follows. First
from every link $ij$, we have the corresponding $P_{ij}$ commuting
with every term in the hamiltonian (\ref{eq:hwen}). The result is
that arbitrary (product) strings of $P$ operators commute provided
they have no ends in common, including closed strings which have no
ends. Secondly, for every even coordinated site we have hermitian
$\Gamma_{k}=i^{(z+2)/2}\prod_{j=-1}^{z}\gamma_{j}(k)$ anticommuting
with every $\gamma$ on that site, and hence commuting with every
term in the hamiltonian; however $\Gamma_{k}$ anticommutes with any
$P$-string ending on site $k$. The maximal commuting set of all
these operators then appears to be the union of:

\noindent (a) all mutually inequivalent closed loop $P$-strings,
equivalent to a minimal set of independent loop terms in the hamiltonian
(\ref{eq:hwen}), numbering $A-N+1$ ;

\noindent (b) the $\Gamma_{k}$ for all even sites, numbering $N_{E}$;

\noindent (c) open $P$-strings ending on odd sites, with no ends
in common and inequivalent under products with closed loop strings,
all of which enumerate to $N_{O}/2$ by taking taking all the odd
sites in (arbitrary) disjoint pairs.

\noindent All of the above commute with each other and with the Hamiltonian,
so we have in total $A-\frac{1}{2}N_{O}+1$ explicit constants of
the motion. Each corresponding operator has eigenvalues $\pm1$, so
its conservation removes one qbit from the dynamics. The author conjectures
that the one extra conservation law relates to conservation of fermion
number (modulo $2$).

Can one exploit the constants of the motion to reduce the size of
the (quantum-mechanically coupled) Hilbert space? Trying this with
$P_{ij}$ operators induces successively less local anticommutation
relations, tending to rebuild the original fermionic representation.
However we can eliminate the site-wise local $\Gamma_{i}$ on even
sites, leading us to a dimension-independent generalisation of what
in one dimension corresponds to the Anisotropic Heisenberg Model representation
of fermions.

Let us focus on some particular even site $i$ and in the following
drop reference to that index. We can eliminate $\gamma_{-1}=i^{(z+2)/2}\gamma_{0}\gamma_{1}..\gamma_{z}\Gamma$,
and then the operators in the Hamilonian which contained $\gamma_{-1}$
take the forms $n=-\frac{i^{z/2}}{2}\gamma_{1}..\gamma_{z}\Gamma+1/2$,
$b_{k}=\frac{1}{2}\gamma_{k'}\gamma_{0}\left(1-i^{z/2}\gamma_{1}..\gamma_{z}\Gamma\right)$,
$b_{k}^{+}=\frac{1}{2}\gamma_{0}\gamma_{k'}\left(1+i^{z/2}\gamma_{1}..\gamma_{z}\Gamma\right)$.
Then because $\Gamma$ still (evidently) commutes with the hamiltonian
we can focus on the sector with eigenvalue $\Gamma=1$ and make this
replacement. Now the only matrices appearing are $\gamma_{k}$, $k=0...z$,
and we can take a minimal anticommuting representation of these\cite{minimalnote},
in terms of which we obtain $n=\frac{1}{2}\left(1+\gamma_{0}\right)$,
$b_{k}=\gamma_{k'}n$, $b_{k}^{+}=n\gamma_{k'}=\gamma_{k'}\left(1-n\right)$
at each even coordinated site.

For an arbitrary graph of even coordinated sites, we now have hamiltonian\begin{multline}
H_{even}=\sum_{i}V_{i}n_{i}+\sum_{<ij>}n_{i}\gamma_{j'}(i)u_{ij}\gamma_{i'}(j)n_{j}\\
+\sum_{<ij..z>}g_{ij..z}\, P_{ij}P_{j.}..P_{.z}P_{zi},\label{eq:heven}\end{multline}
 and the explicit form for the $P_{ij}$ remains as per eqs. (\ref{eq:Pdef}),
except that the dimension of the local Dirac matrices has been halved.
It is now particularly clear how the hopping terms conserve fermion
numbers, and indeed one can further show that $n_{i}\gamma_{j'}(i)u_{ij}\gamma_{i'}(j)n_{j}=n_{i}\left(1-n_{j}\right)\gamma_{j'}(i)u_{ij}\gamma_{i'}(j)\left(1-n_{i}\right)n_{j}$
making the interconnection between occupation numbers $1_{j}0_{i}$
and $0_{j}1_{i}$ totally explicit. The loop terms are also particle
conserving and each $P_{ij}$ factor can be reorganised in similar
manner. If one specialises to $z=2$ corresponding to a one dimensional
chain of sites, the Dirac matrices reduce to Pauli matrices and the
first two terms of the hamiltonian are exactly equivalent to the Anisotropic
Heisenberg spin chain\cite{heisenberg}, well known to represent fermions
in one dimension. The periodic case has one loop term, but this reduces
to a fixed scalar. It is gratifying that such a long-known fermion
hamiltonian turns out to have natural extension to arbitrary dimensions
and even to arbitrary graphs (of even coordination number).

All of our analysis can be generalised to the case where we start
from hopping of an arbitrary number $h$ of half-fermions on each
site. Hitherto we started from one standard fermion per site in the
fermion hopping hamiltonian (\ref{eq:hhop}) and split that into two
half fermions (\ref{eq:decomp}), $h=2$. In the general case any
natural number $h$ is allowed, with the onsite potentials and hopping
matrix elements making arbitrary (hermitian) mixings amongst the $h$
components. For $h=1$ there are no on-site potential terms, and hermiticity
restricts the hopping matrix elements to be pure real, $u_{ij}=u'_{ij}$,
giving us\begin{equation}
H_{hop,1}=i\sum_{<ij>}u'_{ij}m_{0i}m_{0j}.\label{eq:majoranahop}\end{equation}
The standard fermion case discussed ealier is $h=2$, and it reduces
to the sum of two (commuting) $h=1$ hamiltonians if the onsite potential
terms are zero and the $u_{ij}$ are all pure real (to within a gauge
transformation). The case $h=3$ commands interest for particle physics.

The discussion of conservation laws and reduction of Hilbert space
generalises quite trivially, with the understanding that \emph{even}
and \emph{odd} sites are identified by whether $h+z$ is even or odd.
Particularly simple spin models are obtained from $h=1$, where the
local operators $b_{ij}$ are hermitian and can be directly represented
as single local Dirac matrices $\gamma_{j}(i)$ (rather than bilinear
products) requiring only $z_{i}$ Dirac matrices at each site, and
with $B_{i,jk}\propto i\gamma_{j}(i)\gamma_{k}(i)$. Quite general
Majorana fermion hopping hamiltonians can therefore emerge from the
excitations of models built out of simple local spin operators. The
gauge symmetry changes naturally with $h$. For $h=1$ we have only
$Z2$ (sign changes), corresponding to the transformations exploited
in the `KLW trick' of introducing the $S$ operators, and for $h=2$
we already noted $U(1)=O(2)$ gauge symmetry. For $h=3$ we would
have gauge symmetry $O(3)$ if the onsite terms are sufficiently symmetric,
but quite arbitrary mixing of components allowed in the onsite terms
of the hamiltonian would in general reduce the gauge symmetry to $O(2)$.

The simple (hyper-)cubic lattices are of special interest as a source
of insight into the continuum limit, and particle physics interest
focuses particularly on the case where the fermion excitations are
(almost) massless. The special case of the main analysis above previously
presented by Levin and Wen, which was for simple square and cubic
lattices with $h=2$, falls into this category. However that work
also suffered from the standard `Fermion Doubling Problem' of lattice
fermion models\cite{FermionDoubling}, yielding $2^{d}$ massless
fermions in $d$ dimensions.

Starting from a single half-fermion ($h=1$) hopping model halves
the Fermion Doubling to give just $2^{d-1}$ massless fermions in
$d$ dimensions. From the point of view of the spin models, this is
just as natural a starting point as $h=2$. For simplicity we present
the calculation in one dimension, which suffices to demonstrate the
novelty, as the elaboration to higher dimensions simply follows in
the same manner as Levin \& Wen\cite{Wen1,Wenlong}. A simplest one
dimensional Majorana hopping hamiltonian is\[
H_{1}=-iu'\sum_{n}m_{n}m_{n+1}\]
and imposing periodic boundary conditions for such a system of $N$
sites results in a spectrum of standard massless fermions,\[
H_{1}=2u'\sum_{k=0}^{\pi}\left(2\,\widetilde{c}_{k}^{+}\widetilde{c}_{k}-1\right)\sin k.\]
 Here the wavevectors span \emph{half} the first Brillouin zone of
the lattice, $k=n2\pi/N$, $n=1$ to $N/2$, the fermion operators
are given by $\tilde{c}_{k}=1/\sqrt{2N}\sum_{n}e^{-ikn}m_{n}$ and
the negative wavevector Fourier components give their conjugates.
These $\tilde{c}_{k}$ are standard full fermion operators, in particular
obeying $\left\{ \tilde{c}_{k}^{+},\,\tilde{c}_{l}\right\} =\delta_{kl}$.

The above is in effect a staggered fermion\cite{staggered} solution
to the Fermion Doubling Problem\cite{FermionDoubling}. The full fermions
obtained by halving the Brillouin zone can be mapped onto one full
fermion per two original sites, but these are non-locally related
to the original half fermions. For example associating the full fermions
with even sites gives\[
c_{2n}=\sqrt{\frac{2}{N}}\sum_{k=0}^{\pi}e^{ik2n}\tilde{c}_{k}=\frac{m_{2n}}{2}+\frac{i}{\pi}\sum_{q}\frac{m_{2n+(2q+1)}}{2q+1}.\]
This parallels another solution to the Fermion Doubling Problem, in
which highly non-local hopping with amplitudes similar to the above
is used to force a better spectral approximation to the continuum\cite{SLAC}.
Here we have local hopping for the Majorana particles and the non-local
amplitudes encoded in how they map onto conserved full fermions. We
are not forced to work with these non-local realtionships explicitly,
even in the tight binding hamiltonian, because we can always make
full fermions more locally, for example out of adjacent pairs of half
fermions; the price of doing this is that in terms of the latter the
hamiltonian then contains pair creation and annihilation terms.

One can introduce mass without doubling the spectrum, simply by modulating
the bond strengths $u'_{n,n+1}\rightarrow\sqrt{u'^{2}+\epsilon^{2}}+\epsilon(-1)^{n}$
to open up an energy gap at $E=0$ (corresponding to $k=0,\pi$).
One then finds dispersion relation$E(k)=\pm4\sqrt{u'^{2}\sin^{2}k+\epsilon^{2}}$
with two states at each wavevector over the quarter Brilliouin zone
$0\le k<\pi/2$. Adding more cartesian dimensions with suitably frustrated
signs to the couplings leads (as per Levin and Wen\cite{Wen1,Wenlong}
for the full fermion case) to the natural generalisations in higher
dimensions, for example $E(k_{x},k_{y},k_{z})=\pm4\sqrt{u_{x}'^{2}\sin^{2}k_{x}+u_{y}'^{2}\sin^{2}k_{y}+u_{z}'^{2}\sin^{2}k_{z}+\epsilon^{2}}$
in three dimensions. Each increase in dimension generates a doubling
of the spectrum, to two species or a two component fermion in $d=2$,
and a four component fermion in $d=3$, matching standard free fermion
fields.

In overall summary, we can now construct hamiltonians in terms of
strictly local operators (in the sense of their commutation relations)
whose excitations correspond to fermion hopping on whatever graph
is desired. If one accepts inter-site hopping as an approximation
to continuum particle dynamics, then this means that almost any fermion
problem can be constructed out of the excitations of what might loosely
have been described as a Bose system. 

The method has been presented for free (non-interacting) fermion systems.
However it is trivially generalisable to the case of an arbitrary
interaction, provided this is a function of the state number operators
(or other bilinear combinations of local fermion variables). In particular
Coulomb interactions between different sites are allowed, as are Hubbard
interactions. Modifying hopping according to occupation numbers, as
in $t-J$ model, is also readily incorporated.

The complete and relatively local enumeration of all the constants
of the motion may have consequences for string-net condensate interpretations\cite{Wenlong,stringnet}.
By introducing terms in the Hamiltonian coupled to all the conserved
quantities, we are free to remove all the degeneracy of even the gauge-equivalent
combinations of $S$ link operators introduced in eq. (\ref{eq:htowen})
from low energy states, in as many ways as the degeneracy introduced.
If string-net condensate states survive this splitting, they have
to be correspondingly degenerate to start with.

The `spin' hamiltonians constructed to give exactly the specified
fermions are somewhat cumbersome. Are there simpler local spin hamiltonians
which still give fermion excitations? The generalised Anisotropic
Heisenberg hamiltonian is one case in point: setting the onsite terms
to zero and removing the number operator factors from the hopping
terms turns out to give the same as starting from a Majorana hopping
hamiltonian (\ref{eq:majoranahop}). Also, we could by deliberate
construction build hamiltonians which are supersymmetric in their
excitation spectrum, just by adding in the desired bosons. Are there
variations on the theme which are more naturally supersymmetric?

\begin{acknowledgments}
The author acknowledges discussions with R. Roemer, J.B. Staunton,
F. Pinski, B. Muzykantskii and D.R. Daniels. The attention to detail
by an anonymous referee led to significant improvements in the presentation. 
\end{acknowledgments}


\begin{thebibliography}{10}
\bibitem{minimalnote}One can indeed verify directly, for example by taking an explicit
representation of the $\gamma_{k}$, $k=0...z$, that all the products
of any \emph{even} number of those matrices (which is all we have)
acting in the sector $\Gamma=1$ do act in this manner. In reducing
the representation we also induce a relation $\gamma_{0}=\pm i^{z/2}\gamma_{1}..\gamma_{z}$
, where the negative choice of sign has been used in the results following.
\end{thebibliography}
\end{document}